Design, Fabrication and Characterization of FeAl-based Metallic-Intermetallic Laminate (MIL) Composites

Haoren Wang,[a] Tyler Harrington,[b] Chaoyi Zhu,[b] and Kenneth S. Vecchio,[a,b]

[a]Department of NanoEngineering, UC San Diego, La Jolla, CA 92093, USA
[b]Materials Science and Engineering Program, UC San Diego, La Jolla, CA 92093, USA

Abstract: FeAl-based MIL composites of various iron alloys were fabricated with an innovative "multiple-thin-foil" configuration and "two-stage reaction" strategy. Alternating stacked metal foils were reactive sintered via SPS at 600°C and 1000°C to grow intermetallics. The "multiple-thin-foil" configuration reduces reaction time, enables local chemical composition control and allows metal/intermetallic combinations, which cannot be produced via the conventional methods. Fe-FeAl, 430SS-FeAl, and 304SS-FeAl MIL composites can be synthesized with desired metallic/intermetallic ratios, where FeAl is the single intermetallic phase present in the composites. Microstructure analysis via SEM, EDS, and EBSD confirms phase identification and reveals the formation of transition layers. The transition layer, which incorporates the composition gradient between the metal (Fe, 430SS or 304SS) and the FeAl intermetallic phase, provides a gradual change in mechanical properties from the metal to intermetallic layers, and further functions as a chemical barrier into which other undesired intermetallics dissolve. Driven by diffusion-controlled growth, grains in the transition layers and FeAl regions exhibit ordered arrangement and sintering textures. Hardness profiles from the metal layer to FeAl region reveal the correlation between local mechanical properties and local chemical compositions. In compression testing, the compressive strength can reach 2.3 GPa with considerable plasticity, establishing the best mechanical properties of any MIL composites synthesized to date.

Key Words: Intermetallics, Metal matrix composites (MMCs), Iron aluminides, Reactive sintering, Metallic-Intermetallic Laminate (MIL) composites

1. Introduction

Metal-intermetallic laminate (MIL) composites are produced via incorporating layers of ductile metals into strong, but brittle intermetallics for optimizing mechanical behaviors. Generally, aluminide-intermetallics possess ordered crystalline structures with high specific modulus and high specific compressive strength, but often very limited plasticity or toughness. Reinforcing these intermetallics with particles, fibers or layers of ductile metals can enhance the toughness [1], making the materials more efficient for structural applications.

MIL composites are typically synthesized via hot pressing alternating stacked metal foils so that intermetallic layers form as the result of interdiffusion and chemical reaction, while an appropriate pressure ensures intimate contact between the sheets of metal foils. The selection of the foil composition and thickness can determine the physical and mechanical properties of the MIL composites so that the specific performance requirements can be fulfilled. The ability to tailor composite microstructure, and the low cost of the initial metallic foils make MIL composites ideal as commercially scalable structural materials, suitable for aerospace applications that require lightweight materials with high specific properties. By tuning the geometry of the initial metal foils, MIL composites can be synthesized into complex shapes, such as rods, tubes or cones, for specific platforms. In addition, multi-functionality can be incorporated into MIL composites, i.e. having the initial foils pre-machined with cavities to provide pathways in the composites for sensors to be embedded for damage detection [2].

Predecessors of MIL composite materials were first synthesized via a solid-state combustion wave in 1989, but little control of microstructure was achievable due to the self-sustaining reaction kinetics [3]. The concept of 'moderated-reactive sintering' leading to the formation of microstructure-controlled MIL composites was systemically introduced in 2001 by Harach and Vecchio for the Ti-Al system [4]. Since then, studies of MIL composites have primarily focused on the Ti-Al system to understand mechanical behaviors. Rohatgi *et al.* [5] investigated the fracture behavior, Adharapurapu *et al.* [6] investigated the fatigue crack resistance, Li *et al.* [7,8] investigated the damage evolution, Cao *et al.* [9] investigated the ballistic performance, and Jiang et al. [10] investigated the dynamic fracture behavior. The effect of metallic/intermetallic ratio was investigated for optimizing the properties [11,12], and other reinforcements, such as ceramic fibers [13], were also introduced to the Ti-Al system to enhance the properties of the MIL composites. However, $Al_3Ti$, which exhibits little plasticity, is the only intermetallic phase formed when reacting Al foils with Ti foils due to extremely slow growth kinetics for other intermetallics. Therefore, MIL composites in the Ti-Al system exhibit limited plasticity, while the lack of available intermetallics limits the ability to tune the properties of the materials. In order to lower the cost and optimize strength and ductility for MIL composites, Ni-Al [14–16] and Fe-Al [17–19] systems, which process ductile intermetallic phases, have begun to attract more attention in recent years.

In the Fe-Al system, the conventional fabrication process occurs between the eutectic temperature of Al-Fe (655°C) and the melting point of Al (660°C) to achieve the fastest reaction rate without macroscale melting. However, previous studies of microstructure evolution in MIL composites in the Fe-Al system [17,19], including pure iron, 430 stainless steel and 304 stainless steel, suggest that the conventional sintering temperature (655°C to 660°C) only generates brittle intermetallics, such as $Fe_2Al_5$, $Fe_4Al_{13}$ and $Cr_2Al_{13}$. On the other hand, studies using Fe-Al diffusion couples [20] confirms the formation of FeAl at 1000°C, which is reported to be a ductile intermetallic phase [21–23]. Although subsequent annealing of pure Fe-Al MIL composites can transform some intermetallics to FeAl, when either 430 or 304 stainless steel is incorporated, other brittle phases, such as $FeAl_2$ and $Cr_5Al_8$, form [18]. In summary, synthesizing MIL composites with only ductile intermetallics, such as FeAl, remains a significant challenge for the field.

In the present study, an innovative fabrication process for MIL composites is proposed to solve the challenge of controlling selective phase formation. MIL composites, where FeAl is the single intermetallic phase, have been successfully synthesized with pure iron or stainless steels. Microstructure assessment of the composites was investigated to evaluate the fabrication process, and growth kinetics were analyzed to understand the design for sintering parameters. Local mechanical behaviors were estimated via nanoindentation and global mechanical properties were measured via compression testing.

2. Experimental

2.1 Material Processing

Foils of commercial pure 1100 aluminum, pure iron (99.5%), 430 stainless steel (430SS, 18 wt% Cr) and 304 stainless steel (304SS, 18 wt% Cr and 8 wt% Ni) were feedstocks to produce MIL composites. The metal foils were first abraded with steel wool pads to remove surface oxides and contaminants, rinsed in acetone with ultrasound cleaning, and then stacked in the configuration shown schematically in Fig. 1(a2). The full layering involves thick Fe/SS foils that are partially retained as remnant metal layers, and ensembles of alternatingly stacked Al-Fe-Al-Fe-Al thin foils, which are intended to transform into the FeAl phase, in between pairs of the thicker Fe/SS foils. For convenience and distinguishing from the conventional 'thick-foil' stacking demonstrated in Fig. 1(a1), this configuration is hereafter termed the "multiple-thin-foil" configuration. As illustrated in Fig. 1(a3), fabrication of the Fe-FeAl MIL composites involved seven layers of 100 μm Al and six layers of 75 μm Fe foils in each "multiple-thin-foil" ensemble, and a total of four such ensembles were incorporated in the sample. These four ensembles were placed between the 500 μm Fe foils, resulting in a total of five layers of the 500 μm Fe foils in the sample. Fabrication of the corresponding 430SS-FeAl or 304SS-FeAl MIL composites, as shown in Fig. 1(a4), replaced the 500 μm Fe foils by a sandwich structure consisting of 100 μm Fe, 300 μm stainless steel and 100 μm Fe (totaling 500 μm in thick metal layers).

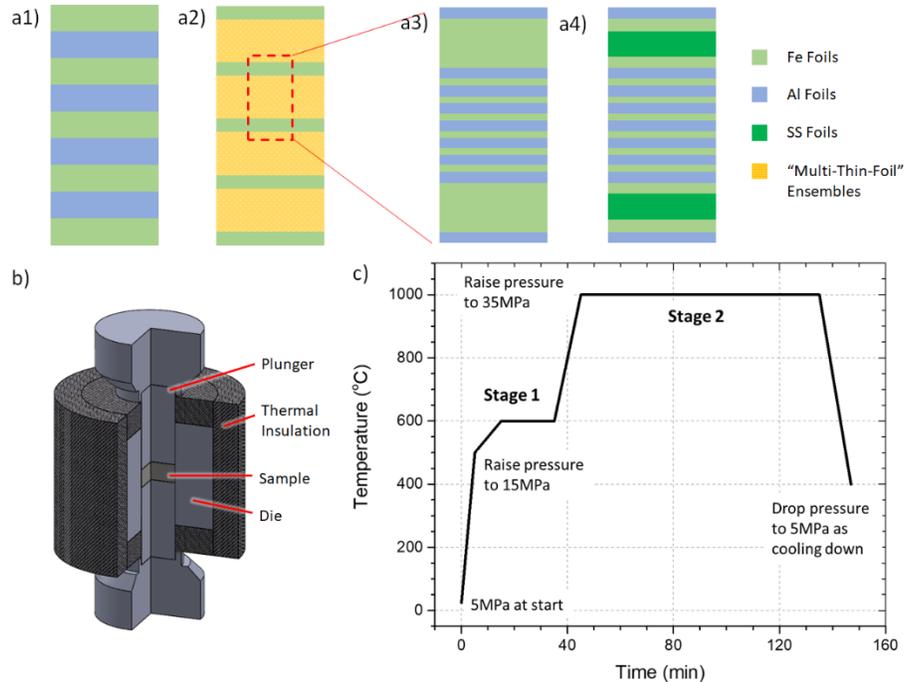

Fig. 1 (a) Schematic diagrams illustrate the metal foils stacks for synthesizing MIL compositions: (a1) Conventional "thick-foil" configuration, (a2) "multiple-thin-foil" configuration invented in this work, (a3) and (a4) details about "multiple-thin-foil" ensembles for pure iron and stainless steel, respectively. (b) An illustration of SPS setup. (c) Sintering parameters profile.

The stacked foils, which were cut into 20mm diameter disks, were placed in a Thermal Technologies Sparking Plasma Sintering (SPS) Model GTAT 10-3 system for reactive sintering. As shown in Fig. 1(b), the SPS assembly consists of a graphite die, with an inner diameter of 20mm, two cylindrical graphite plungers for loading and electric conductivity, and thermal insulation cloth made of carbon fiber. Furthermore, samples were covered by molybdenum foils (99.95%, 0.025-mm-thick) to protect the sample from carbon contamination, and graphite film (0.12-mm-thick) to protect the SPS tooling. The plunger-die assembly is loaded into the vacuum chamber of the SPS machine, loaded axially, and current is passed through the sample via the graphite plungers so that Joule heating brings the sample up in temperature to activate reactive sintering [24]. Fig. 1(c) shows a typical sintering curve for the materials. In stage 1, the temperature is quickly ramped to 500°C, and then slowly ramped to 600°C to prevent over-shooting this target temperature to avoid melting the Al. After holding at 600°C for 20 minutes, the thin Al foils are completely consumed, transforming into $Fe_2Al_5$ phase with the adjacent thin Fe foils. Once all Al is converted to this $Fe_2Al_5$ intermetallic, the temperature can be increased to 1000°C, and held for 1.5 hr as the second stage reaction to form the FeAl phase.

2.2 Characterization

Fe-FeAl, 430SS-FeAl, and 304SS-FeAl MIL composites samples were mounted with layers perpendicular to the surface, and then polished following standard metallographic preparation procedures. The microstructure was investigated using a Thermo-Fisher (formerly FEI) Apreo scanning electron microscope (SEM) equipped with an Oxford Instrument's Energy-Dispersive X-ray Spectrometer (EDS) and an Oxford Instrument's Symmetry electron backscattered diffraction (EBSD) system. SEM images helped to identify metal/intermetallic layers for further EDS and EBSD analysis. EDS line scans were used to measure chemical composition profiles from the metal to intermetallic layers, while EBSD mapping collected crystallographic information for identifying phases and grain orientations (textures). After microstructure characterization, the samples were placed into a KLA (formerly KEYSIGHT) G200 Nanoindenter for measuring hardness profiles across the layers. The tests were performed with the Berkovich tip under a load of 500 mN for 5s and repeated 40 times for each area of interest to ensure statistically representative results.

The sintered MIL composite disks were cut into 6 mm cubes for compression testing following standard ASTM E-9. The cubic geometry allows a straightforward comparison of the mechanical behaviors of the anisotropic composites in different directions. The cube dimension of the specimen were determined based on the samples' thickness, which was nominally 6 mm after sintering. The cubes were ground and polished to remove damage region introduced during cutting and the molybdenum foil that were partially sintered to the top and bottom of the samples. A strain gauge was attached to the specimen, aligned with respect to the loading direction for accurate small-strain measurements; large strain measurements were recorded using crosshead displacement and then corrected with the specimen deformation. Quasi-static compression tests were performed using a standard screw-driven load frame at room temperature with a strain rate of $10^{-3}$/s, and grease was applied to the sample ends to minimize friction. During the test, failure of the specimen was defined as the moment when the load undergoes a significant drop.

3. Results and Discussion

3.1 Microstructure

Fig. 2 shows SEM images of the three MIL composite materials (Fe-FeAl, 430SS-FeAl, and 304SS-FeAl), where the metallic/intermetallic ratio is around 30/70. The intermetallic FeAl phase formed as the result of interdiffusion between alternating stacked thin Fe and Al foils in the "multiple-thin-foil" ensembles, while the remnant metal layers remained from the initial thick metal foils. In contrast to most MIL composites, which exhibit sharp boundaries between metals and intermetallics, in these FeAl-based MIL composites, there exist chemical gradient regions, termed a "transition layer" between the FeAl phase and the pure iron or SS layers, which consists of an $\alpha$-Fe solid solution layer and an FeAl solid solution layer.

3.1.1 Fe-FeAl MIL Composites

Fig. 3(a) shows low magnification SEM micrographs of the microstructures of pure Fe-FeAl MIL composites. The FeAl region is located at the center of the image, while the pure iron layers are on the left and right sides. Considering the translational symmetry of the layering in the MIL composites, examining one region spanning from one remnant metal layer to another can be regarded as representative of the entire sample.

According to Fe-Al phase diagram [25], the phase transformation of body-centered cubic (BCC) $\alpha$-Fe (also known as ferrite) to face-centered cubic (FCC) austenite occurs at 914°C for pure iron. Under the sintering condition of 1000°C, aluminum can dissolve into $\alpha$-Fe phase with a solubility up to 29 at%, while a second order phase transformation (disordered $\alpha$-Fe solid solution to ordered FeAl [B2]) occurs at above 29 at% without a step change in composition. The exact composition threshold for the phase transformation varies slightly with temperature along the $\alpha$-Fe/FeAl solvus. Theoretically, when the temperature drops below 650°C, there exists a compositional step between the $\alpha$-Fe phase and the FeAl phase. Furthermore, when the temperature drops below 545°C, the Fe$_3$Al phase would form prior to the ordered FeAl solid solution. Neither the formation of Fe$_3$Al nor the compositional step between $\alpha$-Fe and FeAl is observed in these FeAl-based MIL composites, which indicates that during the cooling stage, the microstructure evolution that formed at 1000°C remains when cooled to room temperature.

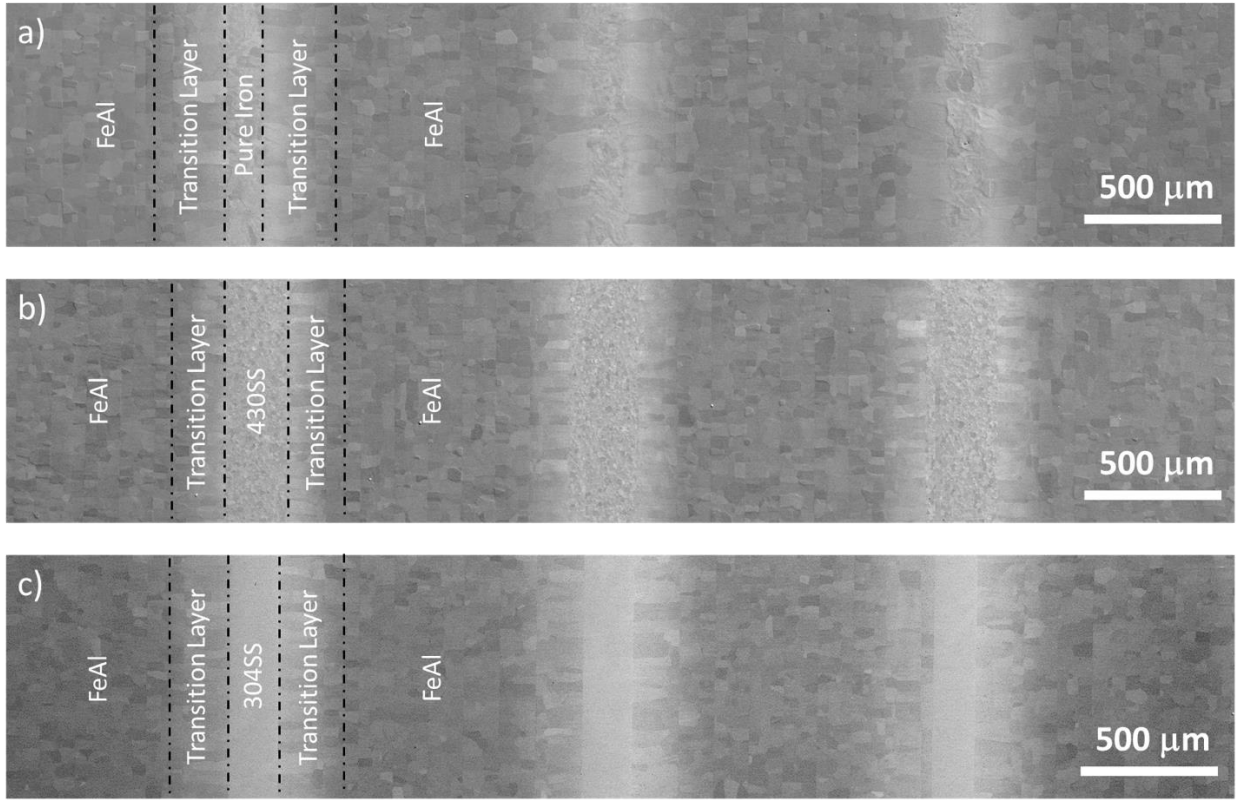

Fig. 2  SEM micrographs showing the structure of (a) Fe-FeAl, (b) 430SS-FeAl and (c) 304SS-FeAl metallic-intermetallic laminate (MIL) composites.  The sintering load direction would be in the horizontal direction perpendicular to the layers.

The thickness of each pure iron layer, measured from Fig. 3(b) and 3(d), which had gone through the ferrite-austenite-ferrite phase transformation, is ~130 μm.   Adjacent to the pure iron layer, an Al-enriched $\alpha$-Fe solid solution layer exists.  Based on the composition profile in Fig. 3(b), the thickness for the $\alpha$-Fe solid solution layer is around 140 μm.  In this work, pure iron layers and $\alpha$-Fe solid solution layers are regarded as the metal portion of the composites.  As revealed in Fig. 3(c), the $\alpha$-Fe solid solution layer is composed of a single layer of columnar grains.  The grain structure indicates that the formation of the α-Fe solid solution layer is driven by aluminum diffusion, and the layer grows epitaxially towards the pure iron as the pure iron layer decreases in thickness.

Adjacent to the α-Fe solid solution layer is a 110μm-thick FeAl solid solution layer. The α-Fe and FeAl phases possess similar crystal structure and lattice constants (bcc and ordered bcc [B2]). Hence, they can exhibit coherent phase boundaries, providing a more gradual change in microstructure. Furthermore, as shown in the red circled region in Fig. 3(c), both α-Fe and FeAl phases can coexist in the same grain. The coherence of the phase boundaries allows dislocations to move more easily across the phase boundary, providing extra ductility and fracture toughness. Typically, the coherent interfaces possess relatively small interfacial energy, but generate strain energy due to lattice mismatch [26]. The lattice mismatch at the α-Fe/FeAl interface is relived as the chemical composition variation across the interface is neglectable, creating similar lattice parameters. The transition layer, referring to the combined α-Fe solid solution layer and the FeAl solid solution layer, can bond the pure iron and FeAl layers together, so that all the layers in the composites can response to the applied stress in a more gradual manner across the layers.

The thickness of the FeAl layer, which contains 48 at% Al, measured from Fig. 3(b) and 3(d), is ~660 μm. The chemical composition of FeAl is slightly off-stoichiometry, partially due to the Fe/Al foils not being the exact thickness ratio to produce stoichiometric FeAl. The non-stoichiometric FeAl phase can exhibit dramatically improved plasticity compared the stoichiometric FeAl [22]. According to the EBSD band contrast map in Fig. 3(c), all of the FeAl grains assemble along the apparent vertical straight lines, which are identified as "centerline" according to previous MIL studies [4]. In the first stage of reactive sintering, as the intermetallics phase grows, contaminants on the original Fe/Al interface are pushed towards the Al foil side, at the transformation interface due to the faster diffusivity of Al. When aluminum is completely reacted, the oxides and impurities pushed from both sides would accumulate at the former aluminum center. Compared to "thick-foil" configuration, "multi-thin-foil" configuration reduces the amount of impurities at centerline, as each centerline concentrates impurities from significant thinner aluminum foil, reducing the impact on mechanical behavior. Centerlines guide the grain alignment as the impurities hinder diffusion and grain growth. Grains grown from the two sides meet at the centerline with minimal curvature, diminishing the driven force for grain growth. Therefore, "centerlines" form within the FeAl grain regions at the middle of the thickness of the thin Al foils of each 'multiple-thin-foil' ensemble.

The grain orientation map and corresponding inverse pole figures shown in Fig. 3(e) and 3(f) reveal the texture of the material. Formation and growth for both FeAl and $\alpha$-Fe solid solution are driven by one-directional diffusion during the reaction, inducing the sintering texture for both phases. The texture aligns towards the <111> direction, rather than <100>, the conventional preferred growth direction, or <110>, the normal direction of close-packed planes for BCC materials. The formation and growth of intermetallics would not only consider the energetically favorable planes, but also be influenced by the orientation of parent grains and the diffusion rate in various directions. The pure iron region would also retain some texture from the Fe foils that were rolled to the desired thickness.

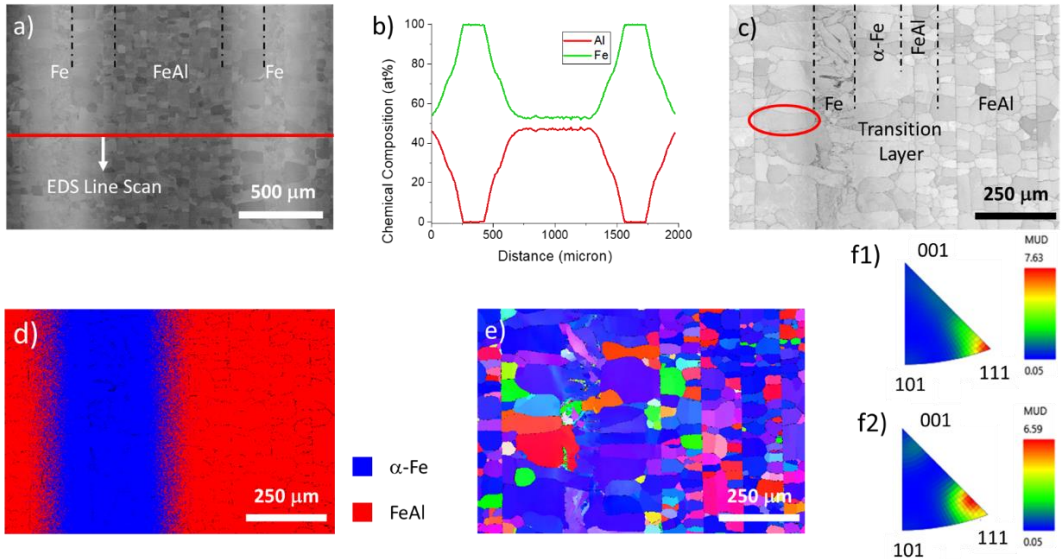

Fig. 3 Microstructure characterization for Fe-FeAl MIL composite. (a) Low magnified SEM image investigates a "repeat unit" of the material. (b) EDS line scan along the red line notated in (a). (c) Band contrast map collected via EBSD reveals the grain structure. Circled in red is a grain where FeAl and $\alpha$-Fe coexist. (d) EBSD phase map for the area in (c) supports phase identification. (e) Grain orientation map along the sintering direction (horizontal) for the region (c) indicates a sintering texture. (f1) and (f2) Inverse pole figures for FeAl and $\alpha$-Fe phase (transition layer), respectively.

3.1.2 430SS-FeAl MIL Composites

Fig. 4(a) shows low magnification SEM micrographs of the microstructures of 430SS-FeAl MIL composites. The FeAl region is located at the center of the image, while the 430SS layers are on the left and right sides.

The thickness of each 430SS layer, measured from Fig. 4(b) and 4(d), is ~220$\mu$m. As illustrated in the experimental section, a pure Fe foil was inserted between the Al and SS foils as a 'buffer' layer to avoid the formation of Cr-Al compounds. 430SS is a ferritic stainless steel, possessing the same α-Fe crystal structure as pure iron with a slightly different lattice constant. The addition of Cr stabilizes the 430SS's α-Fe structure during the sintering. Cr from the 430SS layer diffuses into the pure iron region to reduce the chemical gradient, while Fe also counter-diffuses into the 430SS. The diffusion process reduces the 430SS layer thickness and generates the shoulders of the Fe composition curve in Fig. 4(b) marked by black arrows.

As shown in Fig. 4(c) and 4(f2), the α-Fe solid solution layer adjacent to the 430SS layer is ~100 μm thick, with a similar microstructure and texture as the α-Fe solid solution layer in Fe-FeAl. The penetration of Cr into the α-Fe solid solution terminates within 40 μm into the layer, where the Al concentration is lower than 20 at% (marked by blue arrows). In the α-Fe solid solution region, the Fe concentration is sufficiently high to dissolve all the aluminum and chromium, preventing any Cr-Al intermetallic precipitation. According to previous work [18], without the extra Fe foil between 430SS and Al, the brittle $Cr_5Al_8$ phase would form, which can degrade the ductility and fracture resistance of the material.

The other portions of the material, including the 100-μm-thick FeAl solid solution layer and 670-μm-thick FeAl layer, which are free from Cr, are expected to be identical to the corresponding layers in Fe-FeAl. As revealed in Fig. 4(e) and 4(f1), FeAl grains align by the centerlines with texture along the <111> direction.

3.1.3 304SS-FeAl MIL Composites

Fig. 5(a) shows low magnification SEM micrographs of the microstructures of 304SS-FeAl MIL composites. The FeAl region is located at the center of the image, while the 304SS layers are on the left and right sides.

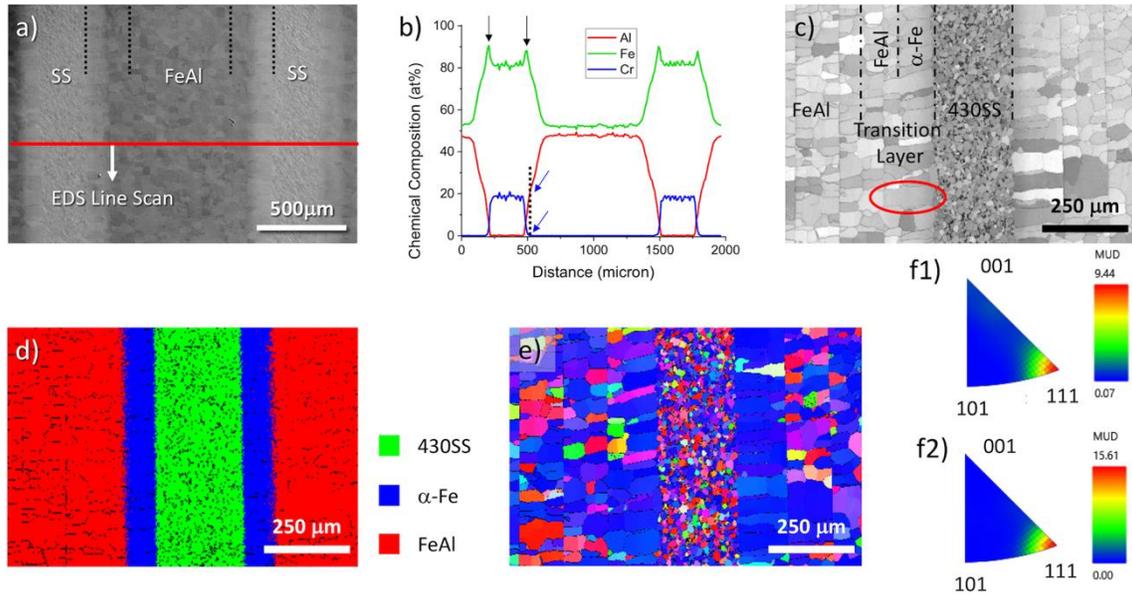

Fig. 4 Microstructure characterization for 430SS-FeAl MIL composite. (a) Low magnified SEM image investigates a "repeat unit" of the material. (b) EDS line scan along the red line notated in (a). (c) Band contrast map collected via EBSD reveals the grain structure. Circled in red is a grain where FeAl and $\alpha$-Fe coexist. (d) EBSD phase map for the area in (c) supports phase identification. (e) Grain orientation map along the sintering direction (horizontal) for the region (c) indicates a sintering texture. (f1) and (f2) Inverse pole figures for FeAl and $\alpha$-Fe phase (transition layer), respectively.

The thickness of each 304SS layer, measured from Fig. 5(b) and 5(d), is ~200 $\mu$m. The 304SS, with an equal amount of Cr as 430SS, is stabilized to the FCC austenite structure at room temperature via the addition of Ni. The loss of Cr and Ni due to diffusion would reduce the 304SS layer thickness, and the region with insufficient Ni would transform into the BCC $\alpha$-Fe structure.

As shown in Fig. 5(c) and 5(f2), the $\alpha$-Fe solid solution layer adjacent to the 304SS layer is ~120 $\mu$m, with the similar microstructure and texture as the $\alpha$-Fe solid solution layer in Fe-FeAl and 430SS-FeAl. Cr and Ni have diffused through the entire $\alpha$-Fe solid solution layer, while the penetration terminated just before entering the FeAl solid solution layer. At the current stage, due to the lack of diffusion data, it is still unclear whether such a phenomenon is just coincident, or the diffusivity of Cr and Ni abruptly drops from $\alpha$-Fe to FeAl. In either case, the $\alpha$-Fe solid solution layer retains a sufficiently high Fe concentration to dissolve all the aluminum, chromium and nickel in solution, preventing the formation of other brittle intermetallics.

As revealed in Fig. 5(e) and 5(f1), the remaining portions of the material, including the 120-$\mu$m-thick FeAl solid solution layer and 670-$\mu$m-thick FeAl layer, which are free from Cr and Ni, are identical to the corresponding layers in Fe-FeAl and 430SS-FeAl. The microstructure of metal and intermetallic layers dissociate from each other, as the metal layers are governed by the initial thick metal foils and the intermetallic layers are governed by the "multiple-thin-foil" ensembles.

3.2 Growth Kinetics and Design of Fabrication Process

In a Fe-Al diffusion couple, $Fe_2Al_5$ would be the first-formed and dominate intermetallic phase at 600°C due to its fast growth kinetics [27]. A very thin layer of $FeAl_3$ may also appear at the aluminum interface [17]. Other Fe-Al intermetallics, including $FeAl_2$, FeAl and Fe$_3$Al, would remain neglectable at 600°C due to extremely slow growth kinetics [27]. Therefore, this study utilizes a "two-stage reaction" strategy to fabricate the FeAl-based MIL composites. The entire purpose for the first stage is to completely consume Al by transforming it to Fe$_2$Al$_5$, so that the samples can subsequently be heated above aluminum's melting temperature enabling more rapid growth kinetics for the FeAl phase.

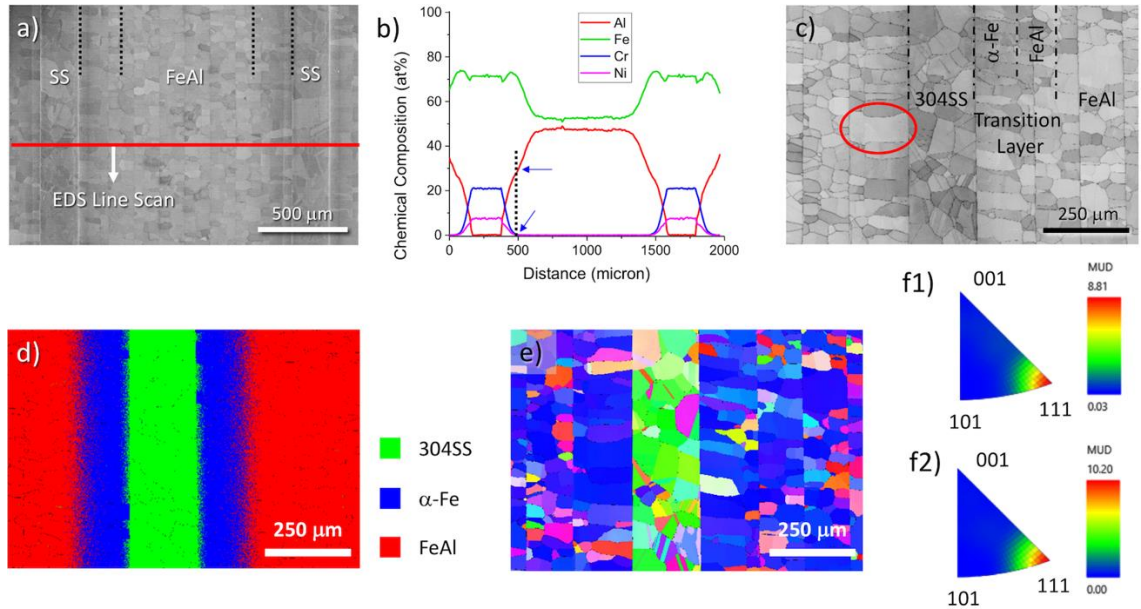

Fig. 5 Microstructure characterization for 304SS-FeAl MIL composite. (a) Low magnified SEM image investigates a "repeat unit" of the material. (b) EDS line scan along the red line notated in (a). (c) Band contrast map collected via EBSD reveals the grain structure. Circled in red is a grain where FeAl and $\alpha$-Fe coexist. (d) EBSD phase map for the area in (c) supports phase identification. (e) Grain orientation map along the sintering direction (horizontal) for the region (c) indicates a sintering texture. (f1) and (f2) Inverse Pole figures for FeAl and $\alpha$-Fe phase (transition layer), respectively.

As the temperature is ramped up in the second stage, diffusion of aluminum would transform Fe$_2$Al$_5$ into FeAl$_2$, and then FeAl. Higher temperatures can accelerate the growth of each phase, saving processing time, but the melting point of Fe$_2$Al$_5$ (1170°C) limits the reaction window. Additionally, since the aluminum diffusivity in α-Fe and FeAl are of similar rates, the formation of FeAl is always accompanied by the α-Fe solid solution.

All the phase transformations discussed above (formation of $Fe_2Al_5$, $FeAl_2$, FeAl and α-Fe solid solution) are diffusion controlled. Reactive diffusion in a binary system can be described by Fick's First Law:

$$J = -D\left(\frac{\partial c}{\partial x}\right) \quad (1)$$

and Fick's Second Law:

$$\frac{\partial c}{\partial t} = D\left(\frac{\partial^2 c}{\partial x^2}\right) \quad (2)$$

In the first stage of the reaction, Al to $Fe_2Al_5$, the system can be treated as a semi-infinite diffusion couple and solved analytically [28]:

$$l^2 = Kt, \quad (3)$$

where $l$ is the intermetallic thickness, $t$ is the time, and the coefficient $K$ is a function of diffusion coefficients and phase boundary compositions. For the subsequent reactions at a higher temperature, the system becomes a finite diffusion couple, so Eq. (3) can no longer predict the thickness of the intermetallics. On the other hand, the system can still be numerical simulated via finite difference analysis by rewriting Eq. (2) in a discrete form:

$$\frac{c_i^{n+1}-c_i^n}{\Delta t} = \frac{D_{i+\frac{1}{2}}^n(c_{i+1}^n-c_i^n)-D_{i-\frac{1}{2}}^n(c_i^n-c_{i-1}^n)}{\Delta x^2}, \quad (4)$$

allowing the required sintering time to be estimated. Meanwhile, both finite and semi-infinite diffusion couples possess self-similarity: if the length scale $x$ is enlarged by $n$, as $X = nx$, extending the reaction time $t$ by $n^2$, as $T = n^2 t$, would generate the same composition profile. The easiest way to understand such self-similarity is to multiple $1/n^2$ to both sides of Eq. (2):

$$\frac{\partial c}{n^2 \partial t} = \frac{\partial c}{\partial (n^2 t)} = \frac{\partial c}{\partial T} = D\left(\frac{\partial^2 c}{n^2 \partial x^2}\right) = D\left(\frac{\partial^2 c}{\partial (nx)^2}\right) = D\left(\frac{\partial^2 c}{\partial X^2}\right). \tag{5}$$

For example, doubling the initial foil thickness would extend the required reaction time fourfold, while reducing the initial foil thickness by 50% reduces processing time by 75%. Therefore, using thinner metal foils as feedstocks could dramatically decrease sintering time and consequently reduce the fabrication cost. Fig. 6 presents the material sintered from conventional alternating stacked 'thick' foils as a comparison group, which utilized multiple 1000μm Fe and 600μm Al foils. In the "multiple-thin-foil" setup, each repeat unit, which refers to one thick metal layer plus one "multiple-thin-foil" ensemble, contains 950μm (500μm+[6*75μm]) Fe and 700μm (7*100μm) Al.

Fig. 6(a) shows the low magnification SEM micrographs of the microstructures for a 'thick-foil' Fe-Al MIL composite fabricated from alternating layers of 1000μm Fe and 600μm Al foils sintered at 600°C for 20 min. Measured from composition profile in Fig. 6(d), the composite consists of 330μm unreacted Al layers, 160μm finger-like $Fe_2Al_5$ layers adjacent to Al and residual pure Fe layers. Predicted from our previous diffusion analysis[17], the additional sintering time $t_a$ required to complete the reaction can be estimated from the equation:

$$\frac{l_0-l}{l_0} = \sqrt{\frac{t}{t+t_a}} \qquad (6)$$

where $l_0$ is the initial Al thickness, and $l$ is the residual Al layer thickness at time $t$. In practice, due to variations in pre-existing surface oxides and uncertainty of exact sample sintering temperature, excess sintering time is often employed to guarantee the completion of the reaction.

Fig. 6(b) shows low magnification SEM micrographs of the microstructure of the 'thick-foil' Fe-Al MIL composite sintered at 1000°C for 1.5 hr, subsequent to the complete consumption of Al. Measured from composition profile in Fig. 6(e), this composite consists of 280μm FeAl$_2$ layers, 350μm residual pure Fe layers and 320μm transition layers of $\alpha$-Fe/FeAl solid solution. Extending the annealing time is necessary to transform the brittle FeAl$_2$ into the ductile FeAl phase of interest. However, in contrast to the 'multiple-thin-foil' Fe-Al MIL composites, where excess sintering brings no additional microstructure changes, in the 'thick-foil' MIL composite the transformation stage from FeAl$_2$ to Fe requires precisely reaction time control to obtain the desired FeAl composition. Fig. 6(c) shows low magnification SEM micrographs of the microstructure of the 'thick-foil' Fe-Al MIL composite sintered at 1000°C for 5 hr. This sample processing condition represents the critical status of the 'thick-foil' Fe-FeAl MIL composites that can be synthesized with the conventional fabrication process. If reaction time is reduced, the reaction will not go to completion, and there will exist a retained brittle FeAl$_2$ layer. In contrast, increasing reaction time will cause the FeAl phase to dissolve into the $\alpha$-Fe solid solution phase, losing both strength and ductility. If thicker iron foils were initially used, the volume ratio of FeAl in the final product would correspondingly decrease, sacrificing strength.

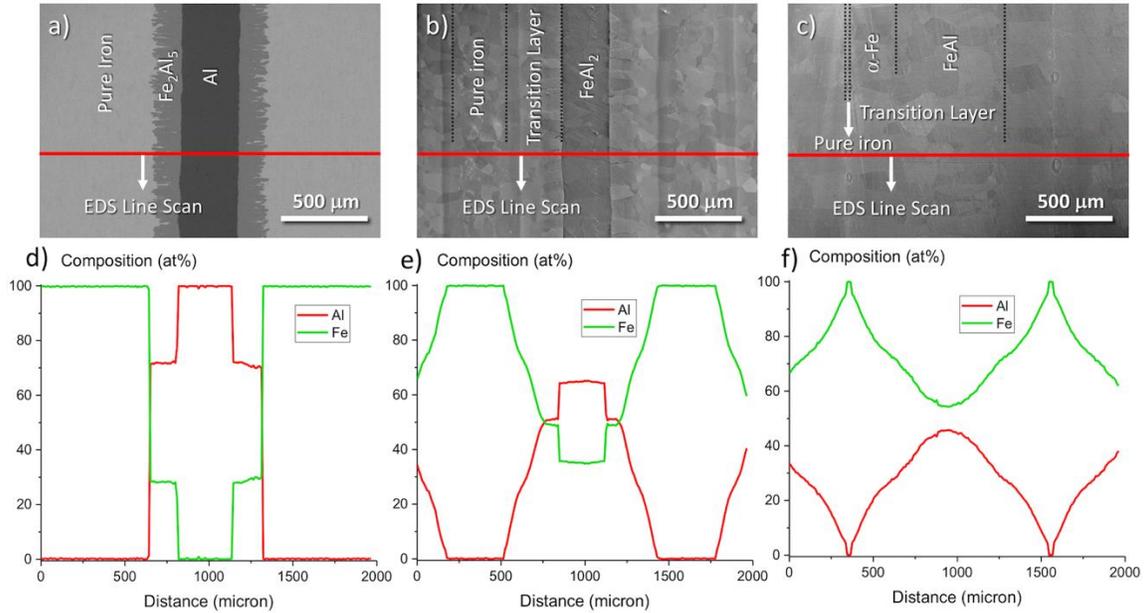

Fig. 6  Microstructure of MIL composites synthesized via conventional alternating stacked 1000μm Fe and 600μm Al foils.  SEM micrographs were collected at the same magnification as Fig. 3(a), 5(a) and 6(a) for straightforward comparison.  (a) Sintered at 600°C for 20 min; (b) Sintered at 600°C for 12 hr, then 1000°C for 1.5 hr; (c) Sintered at 600°C for 12 hr, then 1000°C for 5 hr.  (d), (e) and (f) are corresponding composition profiles from EDS line scans.

On the other hand, since there is already relatively little pure iron remaining with the current Fe/Al ratio, starting with thicker Al foils would end up with the complete consumption of the pure iron layer.  Therefore, the 1000 μm/600 μm foil thickness ratio may represent the only foil combination of the 'thick-foil' configuration that will result in the desired FeAl MIL composite.  Since the initial foil thickness ratio is predetermined by the desired microstructure and diffusion kinetics, the metallic/intermetallic ratio in the fully reacted Fe-FeAl 'thick-foil' MIL composite is fixed to 50:50.  In this case, the major portion of the material is either the $\alpha$-Fe solid solution phase or the FeAl solid solution phase, as demonstrated by the composition curve in Fig. 6(f).

As shown above, the 'thick-foil' configuration is significantly limited due to diffusion kinetics, which restricts both phase type and phase fraction, and the 'multiple-thin-foil' method can overcome these limitations. The ratio of the thin Fe/Al foils independently determines the chemical composition of the intermetallic phase, which in this study is selected to be close to 50:50 (at.%) to form FeAl. Meanwhile, since the thick metal foils utilized in these 'multiple-thin-foil' FeAl MIL composites are retained as the remnant metal layers in the material, their chemical composition, phases, and phase fractions are relatively unchanged during the intermetallic phase formation. As such, almost any metal/intermetallic combination can be achieved with the "multiple-thin-foil" configuration. The thickness of the thin foils determines the required sintering time, and any additional processing time only influences the thickness of transition layers and grain sizes following a parabolic relationship. The number of thin foils in each "multiple-thin-layer" ensemble governs the thickness of the intermetallic layer. By adjusting the number of thin foils and/or selecting the thickness of the thick metal foils, tuning the metallic/intermetallic ratio becomes feasible and convenient. In summary, the "multiple-thin-foil" method has modularized the design for MIL composites, as the intermetallic phase, metal type, grain size and metallic/intermetallic ratio can be independently adjusted to fulfill the specific performance requirements, while both efficiency and robustness for the fabrication process are promoted.

3.3 Mechanical Properties

3.3.1 NanoIndentation Measurements

Fig. 7(a) compares the hardness distribution from metal to intermetallic layers in the Fe-FeAl, 403SS-FeAl and 304SS-FeAl MIL composites, with an example SEM image for the indents on a 304SS-FeAl sample. The reference hardness values for metals were collected from corresponding materials that went through the same heat treatment as MIL composites.

In the hardness profile for the Fe-FeAl MIL composite, the hardness increases with increasing in Al concentration due to enhanced solution strengthening. Based on the theory of solid solution strengthening, the difference in atomic size creates a local stress field, impeding dislocation motion and consequently increasing the hardness. The magnitude of the strengthening effect in a binary system can be estimated based on the equation [23,29–31]:

$$\Delta\sigma = \eta M \mu \varepsilon^{1.5} C^{0.5} \qquad (7)$$

where $\eta$ is an empirical number sensitive to the material [31], $M$ is the Taylor factor which is a constant, $\mu$ is shear modulus, $C$ is the solute concentration and $\varepsilon$ is the misfit parameter, which relates to modulus and atomic size difference. The strengthening effect in intermetallics is more complicated as short-range order, vacancies and other factors need to be taken into consideration [23]. The values collected from the pure iron region of the Fe-FeAl MIL composite are slightly higher than the reference pure Fe foils, but the values for the stainless steel layers in the 304SS-FeAl and 430SS-FeAl MIL composites are quite similar to their respective reference foils.

In the hardness profile for 430SS-FeAl, the hardness gradually rises from the 430SS layer to the FeAl layer. The overall solid solution strengthening effect increases, as the decrease in Cr concentration is overwhelmed by the increase in Al. Studies about Fe-Al-Cr ternary alloys [32] suggest that the addition of Cr to Fe-Al binary system can increase the dislocation line energy via the strengthening of interatomic bonds, and consequently enhances the energy required for dislocation nucleation. Lower dislocation density can enhance the strength at low strain conditions, but is reversed at high strain conditions [33].

In the hardness profile for the 304SS-FeAl MIL composite, the hardness abruptly increases from the 304SS layer to the adjacent transition layer, then gradually drops, before rising again to the FeAl layer hardness. In the 304SS-FeAl MIL composite, Ni diffusion into the Fe transition layer leads to a gradient in hardening from a high value at the 304SS/Fe interface decreasing toward the center of the Fe transition layer, then rises toward the FeAl interface due to the Al gradient hardening the transition layer up to the FeAl hardness.

3.3.2 Compression Tests

The quasi-static stress-strain curves for MIL composites with the layers perpendicular to the load are shown in Fig. 7(c). The compressive strength for Fe-FeAl, 430SS-FeAl and 304SS-FeAl MIL composites are 1250 MPa, 1390 MPa, and 2300 MPa, respectively. The corresponding maximum plastic strain for the three composites are ~0.17, ~0.11 and

~0.17, respectively. During the test for the 304SS-FeAl MIL composite, the experiment was stopped, not due to the sample's failure, but the load limit of the load frame, indicating the actual compressive strength and ductility for 304SS-FeAl would be even higher.

The quasi-static stress-strain curves for these MIL composites with the layers parallel to the load are shown in Fig. 7(d). The compressive strength for Fe-FeAl, 430SS-FeAl and 304SS-FeAl MIL composites are 1750 MPa, 1760 MPa, and 2000 MPa, respectively. The corresponding maximum plastic strain for the three composites are ~0.23, ~0.17 and ~0.19, respectively. As demonstrated in Fig. 7(b2), when the composites are tested with the layers parallel to the load, both metal and intermetallic layers would experience the same amount of stain along the loading direction.

The quasi-static flow curves in Fig. 7(c) and 7(d) for the MIL composites show that the yield strength of the MIL composites is ~700MPa, almost irrespective of the type of metal layer. When loaded perpendicular to the layers, the metal layers will yield first, followed by the gradually yielding of the gradient transition regions. Subsequent to the yield of metal and transition layers, yielding of FeAl determines the apparent yield point for the entire composites, which included work hardening in the metallic layers. When loaded parallel to the layers, since iron and stainless steel possess higher Young's modulus [23] and much lower yield strength than FeAl, the metal layers will also yield first. Subsequently, the local stress applied to the FeAl layers is higher than the global stress readings, and the local stress applied to the metals layers is lower. The strain at the apparent yield point is determined by the FeAl phase, while the stress at the apparent yield point is affeced by the metallic/intermetallic ratio and work hardening of metallic layers.

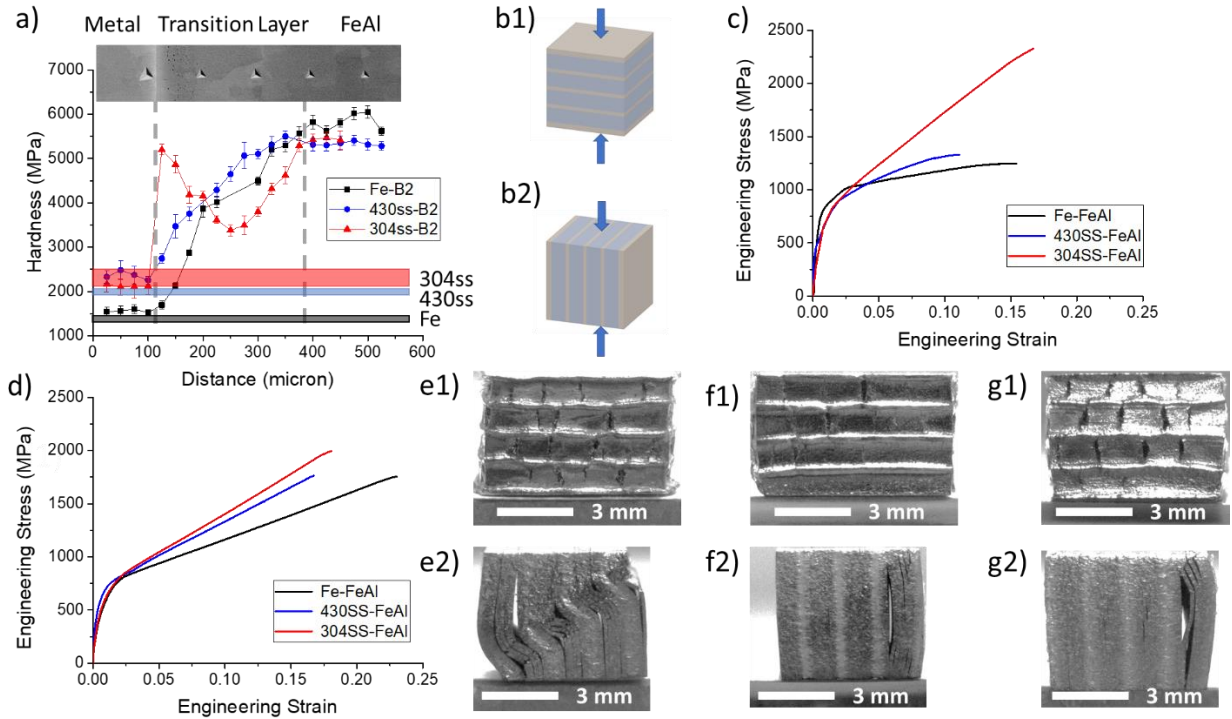

Fig. 7 (a) Hardness curves measured from nanoindentation. The SEM image in the upper part demonstrates typical indent size and distribution. Shaded grey, blue and red bars represent the reference hardness values for the corresponding metals that experienced the same heat treatment. (b1) and (b2) Schematic diagrams illustrate the definition for perpendicular and parallel testing directions, respectively. (c) and (d) Engineering strain-stress cures of the materials compressed in perpendicular and parallel directions, respectively. (e1) and (e2) Photos of Fe-FeAl specimens after compression testing in perpendicular and parallel directions, respectively. (f1) and (f2) Photos of 430SS-FeAl specimens after compression testing in perpendicular and parallel directions, respectively. (g1) and (g2) Photos of 304SS-FeAl specimens after compression testing in perpendicular and parallel directions, respectively.

The quasi-static flow curves also show that both the compressive strength and ductility of the MIL composites depends on the metal layers. Compared to the Fe-FeAl MIL composite, the 430SS-FeAl MIL composite possesses slightly higher strength, but considerably lower plasticity. 430SS is inherently stronger than pure iron, while the addition of Cr also strengthens the transition layers, but at reduced ductility. Although 304SS itself possesses similar mechanical properties with 430SS, the 304SS-FeAl MIL composite exhibits significant improvement in compressive strength and plasticity. The enhancement of the transition layer due to the addition of Ni dramatically enhances the global strain-stress response of the composites.

When MIL composites are loaded parallel to the layers, as shown in Figs. 7(e2), 7(f2) and 7(g2), the layers undergo uniform strains, but significant interfacial stresses evolve as plastic deformation proceeds leading to delamination, after which the separated layers bend and buckle. Rather than the sharp phase boundaries between metals and intermetallics in conventional 'thick-foil' MIL composites, the transition layers formed in the 'multiple-thin-foil' MIL composites enhance bonding between the metal and intermetallic layers, with coherent grain boundaries, and enables an extended strain gradient to develop between the metal layers and intermetallic layers through the transition layers. This effect acts to delay the delamination to much higher strains. When conventional 'thick-foil' MIL composites are loaded perpendicular to the layers, in-plane tensile stresses will induce "axial splitting" failure of the intermetallics, which is a common phenomenon in compression of brittle ceramics. The metal layers adjacent to the fractured intermetallics then become unsupported and fail by shear. Typically, the combination of the two mechanisms generates a macrocrack that propagates along the sample diagonal through the entire thickness [5]. In contrast, when FeAl based MIL composites are loaded perpendicular to the layers, as shown in Figs. 7(e1), 7(f1) and 7(g1), macrocracks only exist in the FeAl region, and never penetrate through the metal layers. Cracks induced by "axial splitting" of the FeAl layer terminate in the transition layers, so shear displacement rarely occur to the metal layers. Further, ambient condition fracture in FeAl typically occurs by transgranular cleavage [22]. As most FeAl grains form columnar grains growing perpendicular to the layering, these FeAl layers possess a high degree of grain shape and crystallographic anisotropy. When loaded parallel to the layers (i.e. perpendicular to the long axes of the columnar grains), the FeAl layers likely exhibit a slightly higher fracture toughness, delaying axial splitting. When loaded perpendicular to the layers (i.e. parallel to the long axes of the columnar grains), this effect appears to be negligible due to the high degree of plasticity in this orientation.

Fig. 8a compares the ductility and compressive strength of the MIL composites, where the data is collected from the strain-stress profiles provided in the corresponding literature [5,15,34-44]. Fig. 8b replots this same data using specific compressive strength in order to compare the results taking density into consideration. By tuning the metallic/intermetallic ratio and intermetallics, MIL composites typically possess either high strength with limited ductility, or relatively high ductility but sacrificing strength. On the other hand, when all these deformation features of the FeAl-based MIL composites are considered together: the enhanced plasticity of the FeAl phase over other aluminide phases, the lack of an obvious and continuous intermetallic centerline, and the presence of a metal to intermetallic transition layer, it is not surprising these FeAl-based MIL composites can exhibit record-high ductility and compressive strength, in both perpendicular and parallel directions, among the known MIL composite family of materials.

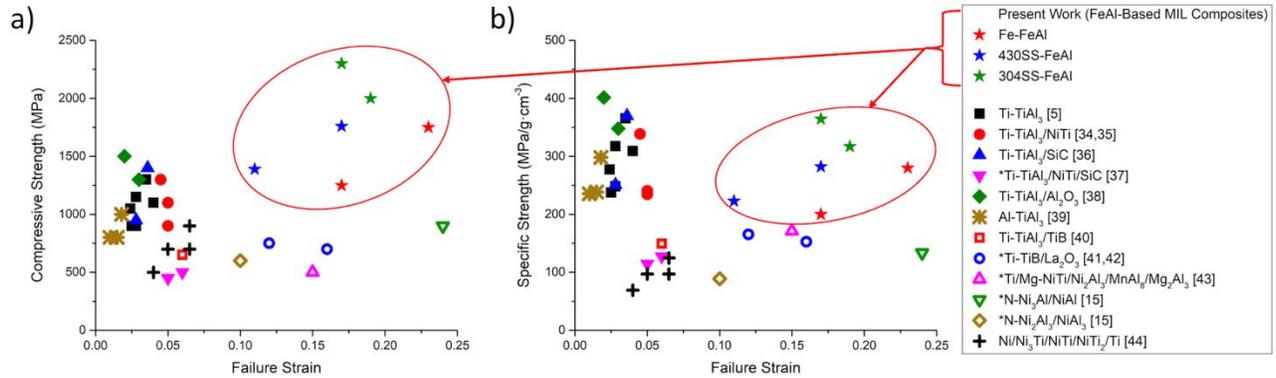

Fig. 8 Comparison of mechanical properties of MIL composites (a) Compressive strength versus strain to failure, (b) Specific strength versus strain to failure. (Ti-TiAl$_3$ [5], Ti-TiAl$_3$/NiTi [34,35], Ti-TiAl$_3$/SiC [36], Ti-TiAl$_3$/NiTi/SiC [37], Ti-TiAl$_3$/Al$_2$O$_3$ [38], Al-TiAl$_3$ [39], Ti-TiAl$_3$/TiB [40], Ti-TiB/La$_2$O$_3$ [41,42], Ti/Mg-NiTi/Ni$_2$Al$_3$/MnAl$_8$/Mg$_2$Al$_3$ [43], Ni-Ni$_3$Al/NiAl [15], Ni-Ni$_2$Al3/NiAl$_3$ [15], Ni/Ni$_3$Ti/NiTi/NiTi$_2$/Ti [44]). *Due to the lack of compression data, the tensile data of the composites is plotted instead. [For (b), if the density was not provided, volume fraction of metallic/intermetallic layers or the starting metal foils were used to estimate the density of the final MIL composites].

4. Conclusion

The present work discussed the material design, fabrication and characterization process to develop a new class of MIL composites. This new class of MIL composites are based on Fe and Al, the two least expensive metals, allowing the potential commercialization of a low-cost series of MIL composites that exhibit excellent mechanical properties. The main conclusions are:

1. Synthesis of MIL composites, where the relatively ductile FeAl is the single intermetallic phase formed, is feasible.

2. The fabrication process for MIL composites is reformed to improve the efficiency and flexibility for material synthesis. The "multiple-thin-foil" configuration saves reaction time, enables local chemical composition control and allows metal/intermetallic combinations, which cannot be produced via the conventional 'thick-foil' methods.

3. Microstructures of Fe-FeAl, 430SS-FeAl and 304SS-FeAl MIL composites were analyzed to evaluate the fabrication process and understand the materials' properties. The transition layer formed between the metal and FeAl regions allows for strain gradients between the metal and intermetallic layers, and additionally functions as a chemical barrier or dissolution layer that limits the formation of other intermetallics.

4. Nanoindentation measurements and compression tests were conducted to estimate both local and global mechanical behaviors. The FeAl-based MIL composites, especially the 304SS-FeAl MIL composite, achieved a new strength and ductility record for the MIL composites family. The mechanisms for the extraordinary mechanical properties are briefly discussed.